\documentclass[final,5p,times,twocolumn,authoryear]{elsarticle}

\usepackage{graphicx}
\usepackage{amsmath}
\usepackage{amssymb, amsmath, amsthm, amsxtra, float, longtable}
\usepackage{color}
\usepackage{natbib}

\usepackage[colorlinks=true,citecolor=blue]{hyperref}

\newcommand{\calv}{C_{\mathrm{A}}}
\newcommand{\cart}{C_{\mathrm{a}}}
\newcommand{\cven}{C_{\bar{\mathrm{v}}}}
\newcommand{\cinh}{C_{\mathrm{I}}}
\newcommand{\cmeas}{C_{\mathrm{meas}}}
\newcommand{\cper}{C_{\mathrm{per}}}
\newcommand{\crpt}{C_{\mathrm{rpt}}}

\newcommand{\lrpt}{\lambda_{\mathrm{b:rpt}}}
\newcommand{\lper}{\lambda_{\mathrm{b:per}}}
\newcommand{\hen}{\lambda_{\mathrm{b:air}}}

\newcommand{\qper}{q_{\mathrm{per}}}
\newcommand{\qalv}{\dot{V}_{\mathrm{A}}}
\newcommand{\qc}{\dot{Q}_{\mathrm{c}}}

\newcommand{\prl}{k_\mathrm{pr}^{\mathrm{rpt}}}
\newcommand{\prm}{k_\mathrm{pr}^{\mathrm{per}}}
\newcommand{\ml}{k_\mathrm{met}^{\mathrm{rpt}}}
\newcommand{\mm}{k_\mathrm{met}^{\mathrm{per}}}

\newcommand{\valv}{\tilde{V}_{\mathrm{A}}}
\newcommand{\vrpt}{\tilde{V}_{\mathrm{rpt}}}
\newcommand{\vper}{\tilde{V}_{\mathrm{per}}}
\newcommand{\vinh}{\tilde{V}_{\mathrm{I}}}

\newcommand{\tb}{\mathbf{p}}
\newcommand{\bb}{\mathbf{b}}
\newcommand{\cb}{\mathbf{c}}

\newcommand{\gb}{\mathbf{g}}
\newcommand{\ub}{\mathbf{u}}
\newcommand{\nulb}{\mathbf{0}}
\newcommand{\yb}{\mathbf{y}}

\newcommand{\di}{\mathrm{d}}

\newcommand{\lk}{\left(}
\newcommand{\rk}{\right)}

\newcommand{\norm}[1]{\lVert #1 \rVert}
\newcommand{\abs}[1]{| #1 |}

\newdefinition{rmk}{Remark}

\journal{Journal of Theoretical Biology}

\begin{document}

\begin{frontmatter}



\title{Physiological modeling of isoprene dynamics in exhaled breath}


\author[label1,label2,label6]{Julian King}
\author[label2,label3]{Helin Koc}
\author[label1,label2]{Karl Unterkofler}
\author[label1,label7]{Pawel Mochalski}
\author[label1]{Alexander Kupferthaler}
\author[label3]{Gerald Teschl}
\author[label4]{Susanne Teschl}
\author[label5]{Hartmann Hinterhuber}
\author[label1,label6]{Anton Amann\corref{cor1}}
\address[label1]{Breath Research Institute, Austrian Academy of Sciences, Rathausplatz 4, A-6850 Dornbirn, Austria}
\address[label2]{Vorarlberg University of Applied Sciences, Hochschulstr.~1, A-6850 Dornbirn, Austria}
\address[label7]{Institute of Nuclear Physics PAN, Radzikowskiego 152, PL-31342 Krak\'{o}w, Poland}
\address[label3]{University of Vienna, Faculty of Mathematics, Nordbergstr.~15, A-1090 Wien, Austria}
\address[label4]{University of Applied Sciences Technikum Wien, H\"ochst\"adtplatz~5, A-1200 Wien, Austria}
\address[label5]{Innsbruck Medical University, Department of Psychiatry, Anichstr.~35, A-6020 Innsbruck, Austria}
\address[label6]{Innsbruck Medical University, Univ.-Clinic for Anesthesia, Anichstr.~35, A-6020 Innsbruck, Austria}

\cortext[cor1]{Corresponding author. Tel.: +43~676~5608520; fax: +43~512~504~6724636.}
\ead{anton.amann@oeaw.ac.at}

\begin{abstract}
Human breath contains a myriad of endogenous volatile organic compounds (VOCs) which are reflective of ongoing metabolic or physiological processes. While research into the diagnostic potential and general medical relevance of these trace gases is conducted on a considerable scale, little focus has been given so far to a sound analysis of the \emph{quantitative} relationships between breath levels and the underlying systemic concentrations. This paper is devoted to a thorough modeling study of the \textit{end-tidal} breath dynamics associated with isoprene, which serves as a paradigmatic example for the class of low-soluble, blood-borne VOCs.

Real-time measurements of exhaled breath under an ergometer challenge reveal characteristic changes of isoprene output in response to variations in ventilation and perfusion. Here, a valid compartmental description of these profiles is developed. By comparison with experimental data it is inferred that the major part of breath isoprene variability during exercise conditions can be attributed to an increased fractional perfusion of potential storage and production sites, leading to higher levels of \emph{mixed} venous blood concentrations at the onset of physical activity. In this context, various lines of supportive evidence for an extrahepatic tissue source of isoprene are presented.

Our model is a first step towards new guidelines for the breath gas analysis of isoprene and is expected to aid further investigations regarding the exhalation, storage, transport and biotransformation processes associated with this important compound.

\end{abstract}

\begin{keyword}
Breath gas analysis \sep Isoprene \sep Volatile organic compounds \sep Modeling \sep Hemodynamics

\PACS 87.80.-y \sep 82.80.Ms \sep 87.19.U

\MSC 92C45 \sep 92C35 
\end{keyword}

\end{frontmatter}


\section{Introduction}
\label{sect:intro}

\subsection{Breath gas analysis and modeling}
Human breath contains a myriad of endogenous volatile organic compounds (VOCs), appearing in the exhalate as a result of normal metabolic activity or pathological disorders. The detection and quantification of these trace gases seems to fulfill all the demands and desires for non-invasive investigation and has been put forward as a versatile tool for medical diagnosis, biomonitoring of disease and physiological function or assessments of body burden in response to medication and environmental exposure~\citep{amannbook,amann2007,amann2004,buszewski2007,rieder2001,miekisch2006,pleil2008}. With the advent of powerful new mass spectrometric techniques over the last 15~years, exhaled breath can nowadays be measured on a breath-by-breath re\-solution, therefore rendering breath gas analysis as an optimal choice for gaining continuous information on the metabolic and physiological state of an individual.

Within the framework sketched above, the success of using VOC breath concentration profiles for tracking endogenous processes will hinge on the availability of adequate physical descriptions for the observable exhalation kinetics of the trace gas under scrutiny. Some major breath constituents have already been investigated in this form, e.g., during exercise conditions or exposure scenarios~\citep{King2010a,moerk2006,anderson2003,kumagai2000,pleilbook}. Nevertheless, VOC modeling remains a challenging task due to the multifaceted impact of physiological parameters (such as cardiac output or breathing patterns~\citep{cope2004}) as well as due to the sparse and often conflicting data regarding potential sources or sinks of such substances in the human body. This paper will be devoted to a thorough study of the \textit{end-tidal} breath dynamics associated with isoprene, which ranks among the most notable compounds studied in the context of breath gas analysis.

\subsection{Isoprene: a survey on physiologically relevant facts}\label{sect:iso}
Isoprene, also known as 2-methyl-1,3-butadiene (CAS number 78-79-5), is an unsaturated hydrocarbon
with a molar mass of 68.11~g/mol and a boiling point of 34$^\circ$C. Isoprene is the most abundant biogenic hydrocarbon emitted by the earth's vegetation and it is also the major hydrocarbon that is endogenously produced by mammals~\citep{gelmont1981}. Its primary source in man has been attributed to the mevalonate pathway of cholesterol biosynthesis~\citep{dsm84}. Originating from acetyl-CoA, mevalonate is transformed into dimethylallyl
pyrophosphate (DMPP). Subsequently, isoprene can be derived from DMPP via an acidic decomposition demonstrated to occur in the cytosol of hepatocytes from rat liver in vitro~\citep{dsm84}. However, whether this final non-enzymatic pathway prevails in the formation of isoprene under physiological conditions continues to be a controversial issue. As has been suggested by several authors, an enzymatic step might catalyze the conversion of DMPP to isoprene in humans~\citep{stone1993,miekisch2004,taucher1997}, similar to the isoprene synthase reaction seen in the chloroplasts of plants and trees~\citep{silver1995}. In this context, possible extrahepatic sites of isoprene production remain to be elucidated. Metabolization of isoprene in mammals primarily rests on epoxidation by cytochrome P450-dependent mono-oxygenases~\citep{monte1985,watson2001}, whereby significant species differences can be observed~\citep{f1996,csanady2001,bogaards2001}. In particular, bioaccumulation in man has been investigated within the framework of toxicological inhalation studies~\citep{f1996}.

Due to its volatility and low affinity for blood (as reflected by a small blood:gas partition coefficient of $\hen = 0.75$ at body temperature~\citep{f1996,kpm01}), isoprene is highly abundant in human breath and accounts for up to 70\% of total hydrocarbon removal via exhalation~\citep{gelmont1981}. Furthermore, it can relatively easily be quantified using a variety of methodo\-logically distinct analytical techniques~\citep{kushch2008,ligor2008,miekisch2006,turner2006,King2010GC}. Apart from being a convenient choice in terms of measurability, breath isoprene has received widespread attention in the literature due to the fact that it may serve as a sensitive, non-invasive indicator for assaying several metabolic effects in the human body (see~\citep{salerno} for an extensive review). 

Most notably, being a by-product of cholesterol biosynthesis as outlined above, breath isoprene has been put forward as an additional diagnostic parameter in the care of patients suffering from lipid metabolism disorders such as hyper\-cholesterolemia. The fact that cholesterol-lowering drugs reduce isoprene output confirms the in vivo relevance of this~\citep{stone1993,kpm01}.~Moreover, interesting relationships between the mevalonate pathway and cell proliferation as well as DNA replication have been discovered~\citep{salerno,rieder2001,fritz2009,brown1980}. Further evidence points toward a strong linkage of breath isoprene levels to different physiological states, thus promoting its general use in biomonitoring, e.g., during sleep or in an intraoperative setting~\citep{amannsleep,cailleux1993,pabst2007}. Despite this huge potential, isoprene breath tests have not yet reached the level of routine clinical methods and are still under development. This is partly due to the fact that drawing reproducible breath samples remains an intricate task that requires further standardization. Furthermore, the decisive mechanisms driving systemic and pulmonary gas exchange are still poorly understood. 

Isoprene concentrations in exhaled human breath exhibit a large variability. In children and adolescents, isoprene excretion in breath appears to increase with age~\citep{taucher1997,smith2010} (with undetectable or very low levels in the breath of neonates~\citep{nelson1998}), until reaching a gender- and age-invariant end-tidal nominal value of about 100~ppb (approx.~4~nmol/l at standard ambient pressure and temperature) characteristic for adults under resting conditions~\citep{kushch2008}. Apart from the factors indicated in the previous paragraph, a number of additional clinical conditions and external influences have been reported to affect isoprene output, including renal dialysis~\citep{capodicasa1999,capodicasa2007,lirk2003}, heart failure~\citep{grath2001}, sleep/sedation~\citep{cailleux1989,amannsleep} and exercise~\citep{kpm01,King2009}. However, the physiological meaning of these changes has not been established in sufficient depth. 

Isoprene can be regarded as the prototype of an exhaled breath VOC exhibiting pronounced rest-to-work transitions in response to physical activity~\citep{kpm01,King2009,turner2006}. We recently demonstrated that end-tidal isoprene abruptly increases at the onset of moderate workload ergometer challenges at 75~W, usually by a factor of about~3--4 compared with the steady state value during rest. This phase is followed by a gradual decline and the development of a new steady state after about 15~min of pedaling~\citep{King2009}, see also Fig.~\ref{fig:compfig}. Since endogenous isoprene synthesis as discussed above has been attributed to pathways with much larger time constants, common sense suggests that the aforementioned rise in isoprene concentration is not due to an increased production rate in the body, but rather stems from changes in hemodynamics or changes in pulmonary function. In this sense, isoprene might also be thought of as a sensitive marker for quantifying fluctuations in blood and respiratory flow. 

With the background material of the previous paragraphs in mind, we view isoprene as a paradigmatic example for the ana\-lysis of low-soluble, blood-borne VOCs, even though it cannot cover the whole spectrum of different physico-chemical characteristics. The emphasis of this paper lies on examining the physiological processes underlying the above-mentioned peak shaped response of \emph{end-tidal} isoprene at the onset of exercise by developing a mechanistic description of the observable exhalation kinetics in normal healthy volunteers. The physical model to be presented here aims at yielding further insights into the flow and distribution route of isoprene in various parts of the human body. Such a quantitative approach is imperative for assessing the relevance and predictive power of extracted breath isoprene concentrations with respect to the endogenous situation and is expected to enhance the fundamental understanding of the physiological role of isoprene in a variety of experimental scenarios.

\section{Experimental basics}
\label{sect:experimental}

\subsection{Setup}
End-tidal isoprene concentration profiles are obtained by means of a \emph{real-time} setup designed for synchronized measurements of exhaled breath VOCs as well as a number of respiratory and hemodynamic parameters. Our instrumentation has successfully been applied for gathering continuous data streams of these quantities during ergometer challenges as well as in a sleep laboratory setting. These investigations aim at evaluating the impact of breathing patterns, cardiac output or blood pressure on the observed breath concentration and permit a thorough study of characteristic changes in isoprene output following variations in ventilation or perfusion. We refer to~\citep{King2009} for an extensive description of the technical details as well as for the various protocols under scrutiny. 

In brief, the core of the mentioned setup consists of a head mask spirometer system allowing for the standardized extraction of arbitrary exhalation segments, which subsequently are directed into a Proton-Transfer-Reaction mass spectrometer (PTR-MS, Ionicon Analytik GmbH, Innsbruck, Austria) for online analysis. This analytical technique has proven to be a sensitive method for the quantification of volatile molecular species $M$ down to the ppb (parts per billion) range by taking advantage of the proton transfer
\[\mathrm{H_3O}^+ + M \to M\mathrm{H}^+ + \mathrm{H_2O}\]
from primary hydronium precursor ions~\citep{lindinger1998,lindinger1998_2}. Note that this ``soft'' chemical ionization scheme is selective to VOCs with proton affinities higher than water (166.5~kcal/mol), thereby precluding the protonation of the bulk composition exhaled air, N$_2$, O$_2$ and CO$_2$. Count rates of the resulting product ions $M\mathrm{H}^+$ or fragments thereof appearing at specified mass-to-charge ratios $m/z$ can subsequently be converted to absolute concentrations of the compound under scrutiny. Specifically, protonated isoprene is detected in PTR-MS at $m/z=69$ and can be measured with breath-by-breath resolution. For further details regarding quantification and the underlying PTR-MS settings used the interested reader is referred to~\citep{schwarzfrag} and~\citep{King2009}, respectively. From the viewpoint of quality control, isoprene time profiles obtained with the setup described above have recently been cross-validated by means of manually extracted GC--MS samples (using solid phase micro-extraction as a pre-concentration step)~\citep{King2010GC}.
Table~\ref{table:measparams} summarizes the measured variables relevant for this paper. In general, breath concentrations will always refer to end-tidal levels. An underlying sampling interval of 5~s is set for each parameter.

\begin{table}[H]
\centering \footnotesize
\begin{tabular}{|lcr|}\hline
{\large\strut}Variable&Symbol&\hspace{-0.25cm}Nominal value (units)\\ \hline \hline 
{\large\strut}Cardiac output &$\qc$ & \hspace{-0.25cm} 6 (l/min)~\citep{mohrman2006}\\
{\large\strut}Alveolar ventilation &$\qalv$ &\hspace{-0.25cm} 5.2 (l/min)~\citep{westbook}\\
{\large\strut}Isoprene concentration &$\cmeas$ &\hspace{-0.25cm} 4 (nmol/l)~\citep{kushch2008}\\
\hline
\end{tabular}
\caption{Summary of measured parameters together with some nominal values during rest, assuming ambient conditions; breath concentrations refer to end-tidal levels.}\label{table:measparams}
\end{table}

\subsection{Recent results and heuristics}\label{sect:res}
This section serves to collect some experimental evidence supporting the hypothesis of a peripheral tissue source of isoprene formation in man, derived from dynamic breath concentration measurements under exercise conditions. The rationale given here mainly builds on our earlier phenomenological studies in~\citep{King2009} and~\citep{King2010GC}. Complementary experiments will be indicated where appropriate. All results are obtained in conformity with the Declaration of Helsinki and with the necessary approvals by the Ethics Commission of Innsbruck Medical University.

\begin{figure}[H]
\centering
\begin{tabular}{c}
\includegraphics[width=9cm]{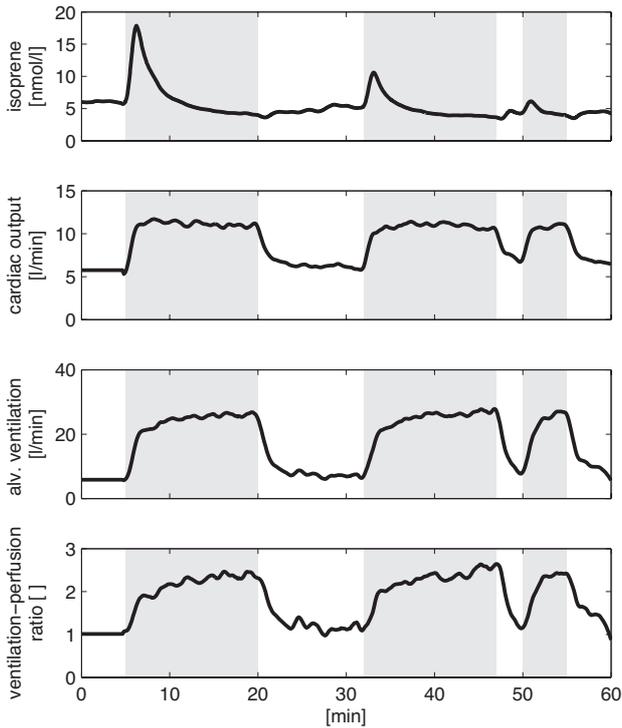}
\end{tabular}
\vspace{-0.5cm}
\caption{Typical smoothed profiles of end-exhaled isoprene concentrations and physiological parameters in response to two-legged ergometer exercise at 75~W. Data are taken from~\citep{King2009} and correspond to one \emph{single} healthy male volunteer (26~years, 72~kg bodyweight). Workload segments are shaded in grey.}\label{fig:compfig}
\end{figure}

Investigating an ensemble of eight normal healthy volunteers,~\citet{King2009} recently demonstrated that isoprene evolution in end-tidal breath exhibits a very reproducible and consistent behavior during moderate exercise scenarios. For perspective, Fig.~\ref{fig:compfig} shows typical results corresponding to a bicycle ergometer challenge of one \emph{single} volunteer under a constant workload of 75~W with several periods of rest.

Generally, starting from a steady state value of about 4~nmol/l during rest, isoprene concentrations in end-tidal air exert a pronounced peak at the onset of exercise (corresponding to an increase by a factor of up to~4). This phase is followed by a gradual decline and the development of a new steady state after approximately 15~min of pedaling. Interestingly, by repeating this regime, the peak size after intermediate exercise breaks can be demonstrated to depend on the duration of the resting phase, despite almost identical profiles of cardiac output and alveolar ventilation. Full recovery of the initial height requires about one hour of rest. A valid model for the description of isoprene concentrations in end-tidal air should be able to faithfully reproduce this wash-out behavior.\\

The aforementioned peak shaped behavior of isoprene has mainly been attributed to its low blood:gas partition coefficient $\hen = 0.75$. According to classical pulmonary inert gas elimination theory (cf.~\ref{app:farhi}), the low affinity for blood implies a high sensitivity of the associated breath concentrations with respect to changes in ventilation or perfusion. More specifically, the basic Farhi equation~\eqref{eq:farhi} predicts that, other factors being equal, increasing/decreasing the alveolar ventilation will decrease/increase exhaled breath concentrations (due to increased/decreased dilution), whereas the relationship between breath concentrations and cardiac output is monotonic and reflects dependence on supply. Using similar reasoning,~\citet{kpm01} proposed a simple quantitative description of breath isoprene concentration time courses during exercise, which is now widely accepted as ``standard model''. However, as has already been argued in~\citep{King2009}, their formulation is deficient in several regards. A principal criticism is that the model of Karl~et~al. essentially relies on a markedly delayed rise of alveolar ventilation with respect to pulmonary blood flow, a premise which clearly contrasts experimental evidence (see, e.g., Fig.~\ref{fig:compfig} as well as~\citep{w92,lumbbook}). The onset of the ventilatory response to exercise is instantaneous and may actually precede the latter (possibly being part of a learned response), so a delay as required above is highly unlikely. Consequently, when subjecting this model to real data streams including measured profiles of pulmonary blood flow $\dot{Q}_{c}$ and alveolar flow $\dot{V}_{A}$, it fails to capture the observed isoprene data, see Fig.~\ref{fig:fit}. 

Further insights into the decisive components affecting breath isoprene excretion can be gained by comparing its dynamic behavior with the profiles of blood-borne VOCs expected to show similar exhalation kinetics. In this context, it has recently been pointed out that breath concentrations of endogenous butane (considered to originate from protein oxidation and/or bacteria production in the colon~\citep{kharitonov2002}) during ergometer exercise resemble the trend anticipated from Equation~\eqref{eq:farhi}, while isoprene exhibits an entirely different qualitative response~\citep{King2010GC}. This is certainly counter-intuitive, as butane is widely comparable with isoprene in terms of various functional factors expected to affect pulmonary gas exchange (including, e.g., blood and tissue solubility as well as molecular weight). 

In light of this discrepancy, it can be conjectured that some unknown substance-specific (release) mechanism has to be taken into account for capturing the exhalation dynamics of isoprene. In order to restrict the number of potential tissue sources for this effect, in a series of auxiliary experiments the ergometer protocol sketched above was modified as follows. Instead of pedaling with both legs, we orchestrated several one-legged workload challenges on a standard ergometer, alternating bet\-ween left and right limb for doing the exercise. The heel of the non-working leg rested on a small chair placed beside the bicycle. Special care was taken to ensure a comfortable seating position of the volunteer so that any contractive movement of the resting leg for stabilization purposes could be avoided. The hand rest of the ergometer was adjusted in such a way that the test subjects could maintain their torso in an upright position throughout the measurement period, with both arms stretched. A constant resistance of 50~W was imposed for the entire experiment and pedaling cadence was maintained at 60~rpm. 

\begin{figure}[H]
\centering
\begin{tabular}{c}
\includegraphics[width=9cm]{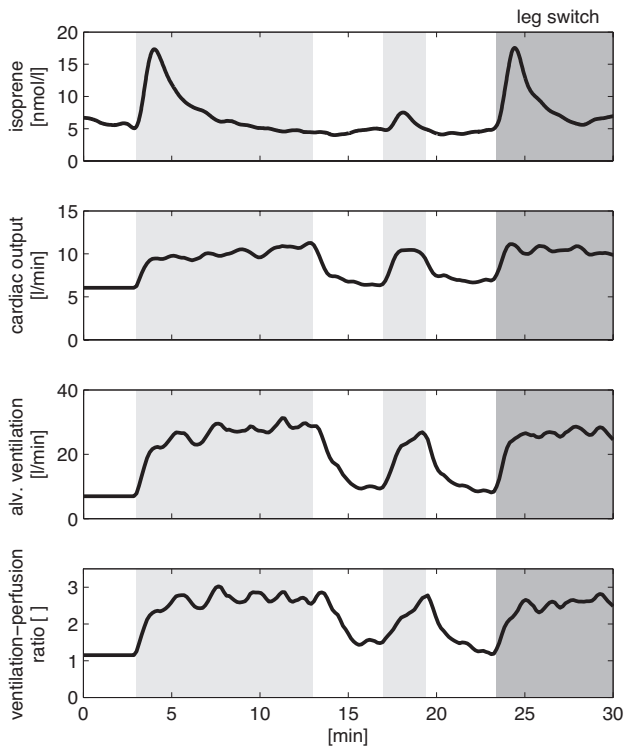}
\end{tabular}
\vspace{-0.5cm}
\caption{Typical smoothed profiles of end-exhaled isoprene concentrations and physiological parameters in response to one-legged ergometer exercise at 50~W. Data correspond to one \emph{single} healthy male volunteer (27~years, 75~kg bodyweight). Left and right leg exercise segments are shaded in light and dark grey, respectively.}\label{fig:compfig2}
\end{figure}

In total, five normal healthy volunteers (age 27-34~years, 4~male, 1~female) were recruited and investigated in this way. No test subject reported any prescribed medication or drug intake. No special restrictions regarding pre-experimental food intake were applied, as this variable seems to have a negligible effect on breath isoprene concentrations~\citep{smith1999,kino2008}. However, volunteers were required to rest at least 20~minutes prior to analysis due to the significant impact of physical activity as discussed in Section~\ref{sect:iso}. Within this time informed consent was obtained regarding the experimental protocol. Additional instrumentation and monitoring closely followed the general procedure reported in~\citep{King2009}. Fig.~\ref{fig:compfig2} shows a representative experimental outcome for one \emph{single} volunteer. 

At the beginning, the qualitative response of end-tidal isoprene concentrations closely resembles the situation presented in Fig.~\ref{fig:compfig} for the two-legged case. After 10~minutes of pedaling with the left leg, followed by a resting period of 4~minutes, a clear wash-out effect becomes discernible, yielding a significantly lower peak height when continuing the exercise with the same leg. However, if the working limb is now switched to the \emph{right} leg (after an intermediate break of 4~minutes as before), an almost complete recovery of the initial peak size can be observed (cf. the time frame between 23~and~30 minutes in Fig.~\ref{fig:compfig2}). On the contrary, it should be noted that the associated rise in cardiac output and alveolar ventilation is of comparable order within all three workload phases. These basic characteristics could reliably be reproduced within the entire collective of test subjects. In particular, consistent results are obtained if the leg switch is from right to left. 

Combining the aforementioned findings provides a clear hint that breath isoprene levels during exercise are linked to local variations of gas exchange in peripheral tissue groups. In particular, they open up a new line of supportive evidence for peripheral production sites of isoprene as indicated in Section~\ref{sect:iso}. Furthermore, the common viewpoint that the breath isoprene peaks characteristic for exercise conditions can mainly be traced back to altered pulmonary gas exchange conditions (resulting, for instance, from an impairment of cardiac output and ventilatory drive~\citep{kpm01}) or local generation in the respiratory tree (as in the case of NO release in the paranasal sinuses during humming~\citep{weitzberg2002}) has to be rejected. As will be discussed in the modeling sections below, we attribute the observable wash-out behavior of isoprene to an increased fractional perfusion of potential storage and production sites, leading to higher levels of the \emph{mixed} venous blood concentration at the onset of physical activity. While the exact tissue groups involved in this process remain speculative, possible origins might include the skeletal locomotor muscles themselves but also the walls of the vascular tree, both of which receive a disproportionately high share of blood flow during exercise. 
There are some indications in the literature that isoprene synthesis can play a role at these sites~\citep{miekischblood,brown1980}. However, further biochemical investigations will need to clarify whether an appropriate metabolic pattern exists in these extrahepatic tissues.

%
\section{Isoprene modeling}
\label{sect:isomodel}
%
\subsection{Preliminaries and assumptions}
\label{sect:prelim}

For the sake of maintaining a balance between tractability and sufficient complexity of the model structure, we shall adopt the usual compartmental approach in our attempts to describe the end-tidal isoprene behavior outlined above. This approach consists in dividing the body into an ensemble of roughly homogenous tissue control volumes that are interconnected via the arterial and venous network~\citep{reddybook,leung1991,gerlowski1983,fiserovabook}. Previously developed physiologically based descriptions of isoprene pharmacokinetics in man and rodents can be found in~\citep{f1996,nih1999,melnick2000,bogaards2001}. These mainly centered on quantifying body burden in response to severe environmental exposure (driven by concerns about the carcinogenic potential of isoprene and/or its metabolites~\citep{melnick1994,nih1999}) and hence often neglected the relatively small contribution of endogenous production to overall bioaccumulation. In contrast, here we will mainly focus on the characteristics of isoprene formation and distribution within specific body tissues under normal physiological conditions. Similarly to the models mentioned above, two major aspects of isoprene exchange will be taken into consideration.

\subsubsection{Pulmonary gas exchange}
\label{sect:gasex}

Following a general premise of classical pulmonary inert gas elimination theory (see~\ref{app:farhi}), we postulate that uptake and removal of isoprene takes place exclusively in the alveolar region. In particular, any pre- and post-alveolar absorption and release mechanisms occurring in the conductive airways (e.g., due to interactions with the tracheo-bronchial lining fluid~\citep{anderson2003,anderson2007,King2010a}) are assumed to be negligible, which is a reasonable requirement for low-soluble VOCs such as isoprene~\citep{anderson2003}. The lung function will be taken into account by considering one single homogenous alveolar unit characterized by an averaged ventilation--perfusion ratio close to one during resting conditions. While this approach ignores the regional ventilation--perfusion scatter throughout the lung, it constitutes a convenient simplification that is justified by the need to keep the parameterization as parsimonious as possible at this stage of the modeling phase. Delivery and elimination of isoprene within the alveolar tract will be governed by cardiac output $\qc$ and alveolar ventilation $\qalv$, respectively, thereby neglecting the small intrapulmonary shunt and alveolar dead space fraction~\citep{lumbbook}. Owing to its lipophilic characteristics and small molecular size, isoprene can be assumed to rapidly pass through the alveolar tissue barrier, so that an instantaneous diffusion equilibrium will be established between end-capillary blood and the free gas phase. This is likely to hold true also under moderate, sub-anaerobic exercise conditions~\citep{wagner2008}.
In the absence of chemical bindings with blood it can thus be deduced that the concentration $\cart$ of isoprene in \emph{arterial} blood leaving the lungs is proportional to the concentration $\calv$ within the alveoli, viz.,
\begin{equation}\label{eq:cart}
\cart=\hen\calv.
\end{equation}
Here, $\hen$ denotes the isoprene-specific blood:gas partition coefficient as introduced in Section~\ref{sect:iso}.

\subsubsection{Body compartments}
\label{sect:bodycomp}
The systemic part of the model incorporates two well-mixed functional units: a richly perfused tissue (rpt) compartment, lumping together tissue groups with comparable blood:tissue partition coefficient $\lrpt \approx 0.4$ (viscera, brain, connective muscles, skin), as well as a peripheral tissue compartment, representing an effective buffer volume that acts as a reservoir for the storage of isoprene (tentatively skeletal muscles). Both compartments are separated into an intracellular space and an extracellular space (including the vascular blood and the interstitial space), whereby a venous equilibrium is assumed to hold at these interfaces. The relevant blood:tissue partition coefficients are summarized in Table~\ref{table:param}. Due to the low fractional perfusion of adipose tissue, an extra fat compartment was not considered. 

In order to capture the redistribution of systemic perfusion during bicycle ergometer exercise, fractional blood flow $\qper \in (0,1)$ to peripheral tissue is assumed to resemble fractional blood flow to both legs. The latter increases with cardiac output and will be modeled as
\begin{multline}\label{eq:qper}\qper(\qc):=\qper^{\mathrm{rest}}+(\qper^{\mathrm{max}}-\qper^{\mathrm{rest}})\times \\ \big(1-\exp{(-\tau\,\max\{0,\frac{\qc-\qc^{\mathrm{rest}}}{\qc^{\mathrm{rest}}}\})}\big), \quad \tau>0. \end{multline} 
Reference values for the indicated variables can be found in Table~\ref{table:param}. For perspective, in the sequel we set $\qper^{\mathrm{rest}}= 0.08$ and $\qper^{\mathrm{max}}=0.7$ (which approximately corresponds to the fractional perfusion of both legs during bicycle exercise at 75~W~\citep{sullivan1989}). The constant $\tau$ will be estimated in Section~\ref{sect:simest}. Alternatively, the right-hand side expression in~\eqref{eq:qper} might also be replaced with a piecewise constant function taking values $\qper^{\mathrm{rest}}$ and $\qper^{\mathrm{max}}$ during rest and exercise, respectively.

As has been mentioned previously, the tissues contributing to isoprene formation are not fully established. In view of the biochemical and experimental results in Sections~\ref{sect:iso} and~\ref{sect:res}, respectively, two distinct non-negative production rates $\prl$ and $\prm$ are incorporated into the model. These values quantify potential hepatic and peripheral sources of endogenous isoprene, the latter being interpreted as a by-product of the biosynthesis of polyisoprenoid compounds, their degradation, or both. While isoprene production in general appears to be subject to diurnal variations~\citep{cailleux1989,amannsleep}, within the typical experimental time frame considered here both rates are treated as constant. Analogously, metabolization of isoprene is described by conventional first order kinetics and will be captured by introducing two rate constants $\ml$ and $\mm$, reflecting cytochrome P450 activity in liver and extrahepatic tissues, respectively~\citep{f1996}. Other ways of isoprene clearance such as excretion via the renal system are considered as long-term mechanisms in this context and will thus be ignored. The specific values for the production and metabolization rates introduced above will have to be estimated based on experimental results and may depend on the individual volunteer investigated. The latter case would be particularly interesting in the light of the fact that isoprene may reflect certain aspects of endogenous cholesterol synthesis.

\subsection{Model equations and a priori analysis}
\begin{figure}[H]

\centering
\begin{tabular}{c}
\includegraphics[scale=1]{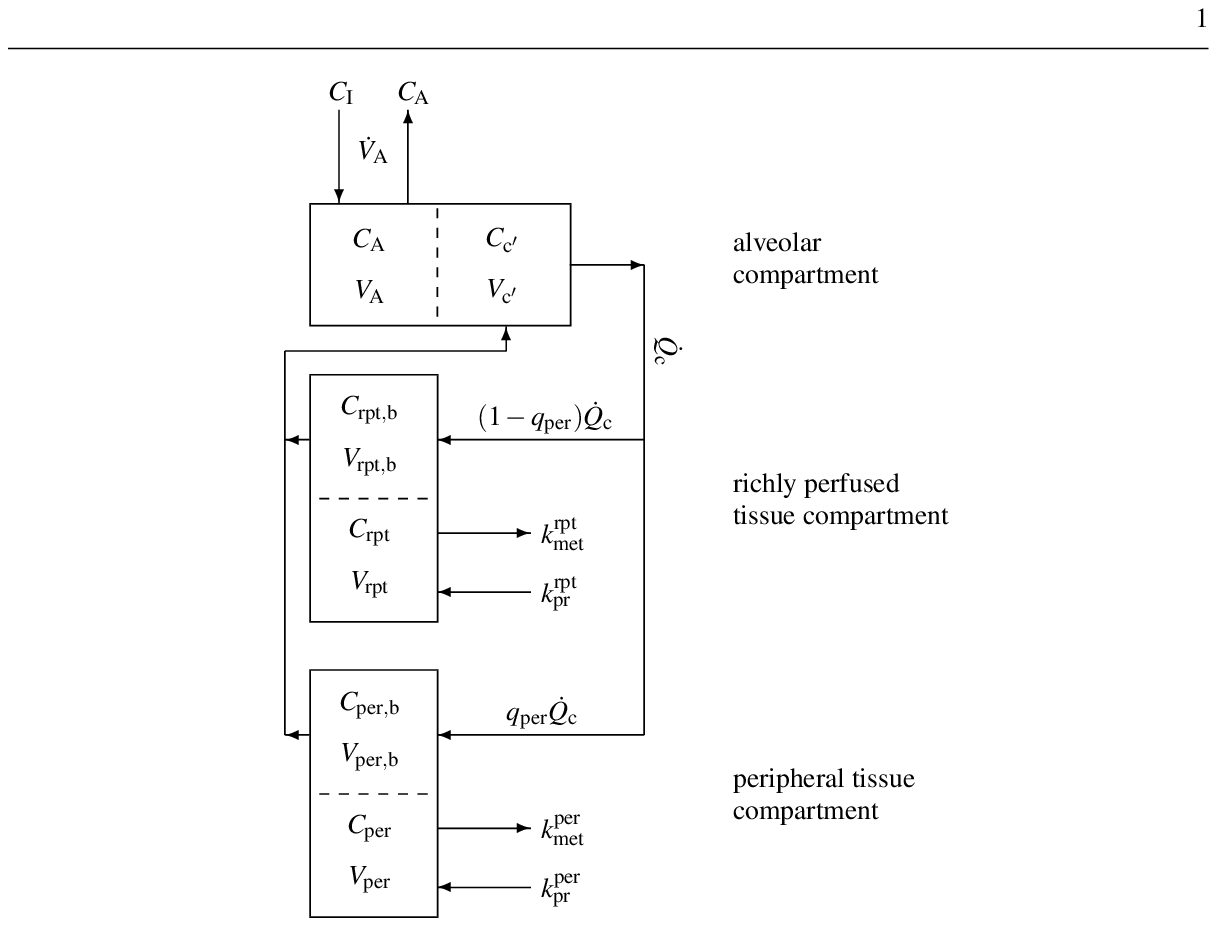}
\end{tabular}
\caption{Sketch of the model structure. The body is divided into three distinct functional units: alveolar/end-capillary compartment (gas exchange), richly perfused tissue (metabolism and production) and peripheral tissue (storage, metabolism and production). Dashed boundaries indicate a diffusion equilibrium. Abbreviations connote as in Table~\ref{table:param}.}\label{fig:model_struct}
\end{figure}

In order to capture the gas exchange and tissue distribution mechanisms presented in the previous paragraphs, the model consists of three different compartments.
A sketch of the model structure is given in Fig.~\ref{fig:model_struct} and will be detailed in the following. Model equations are derived by taking into account standard conservation of mass laws for the individual compartments. In view of the diffusion equilibria postulated in Section~\ref{sect:prelim}, the compartment capacities are governed by the \emph{effective} volumes $\valv:=V_{\mathrm{A}}+V_{\mathrm{c'}}\hen$, $\vrpt:=V_{\mathrm{rpt}}+V_{\mathrm{rpt,b}}\lrpt$ as well as $\vper:=V_{\mathrm{per}}+V_{\mathrm{per,b}}\lper$. Nominal values for the indicated parameters are given in Table~\ref{table:param}.

According to Fig.~\ref{fig:model_struct}, the mass balance equation for the alveolar compartment reads
\begin{equation}\label{eq:alv}
\valv\frac{\di\calv}{\di t}=\qalv(\cinh-\calv)+\qc(\cven-\cart),
\end{equation}
with $\cinh$ denoting the inhaled (ambient) gas concentration, while for the richly perfused and peripheral tissue compartment we find that
\begin{equation}\label{eq:rpt}
\vrpt\frac{\di\crpt}{\di t}=(1-\qper)\qc(\cart -\lrpt\crpt)+\prl-\ml\lrpt\crpt,
\end{equation} 
and
\begin{equation}\label{eq:per}
\vper\frac{\di\cper}{\di t}=\qper\qc(\cart -\lper\cper)+\prm-\mm\lper\cper,
\end{equation}
respectively. Here, the associated concentrations in mixed venous and arterial blood are given by
\begin{equation}\label{eq:cven}\cven:=(1-\qper)\lrpt\crpt+\qper\lper\cper\end{equation}
and Equation~\eqref{eq:cart}, respectively.
Moreover, we state that the measured (end-tidal) isoprene concentration equals the alveolar level, i.e.,
\begin{equation}\label{eq:meas}y:=\cmeas=\calv.\end{equation}
Note that in Equations~\eqref{eq:cart} and~\eqref{eq:cven} it is tacitly assumed that any transport delays between tissues, heart and lung can be neglected. A more refined formulation in this regard can be achieved by considering delay differential equations, see for instance~\citep{bkst2007}.\\

\begin{rmk}
For later purposes, we note that a model accommodating the experimental situation during exhalation and inhalation to and from a fixed volume exposure atmosphere can simply be derived by augmenting Equations~\eqref{eq:alv}--\eqref{eq:per} with an additional compartment obeying
\begin{equation}\label{eq:bag}
\vinh\frac{\di\cinh}{\di t}=\qalv(\calv-\cinh).
\end{equation}
This typically describes closed system (rebreathing) setups such as in~\citep{f1996}.
\end{rmk}

Some fundamental model properties are discussed in~\ref{sect:apriori}. In particular, the components of the state variable $\cb:=(\calv,\crpt,\cper)^T$ remain non-negative, bounded and will approach a globally asymptotically stable equilibrium $\cb^e(\ub)$ once the measurable external inputs $\ub:=(\qalv,\qc,\cinh)$ affecting the system are fixed. This corresponds, e.g., to the situation encountered during rest or constant workload, see Fig.~\ref{fig:compfig}. Analogous results can be established for the augmented system incorporating Equation~\eqref{eq:bag}, describing the evolution of the composite state variable $\underline{\cb}:=(\calv,\crpt,\cper,\cinh)^T$. In this case, the corresponding equilibrium for fixed inputs will be denoted by $\underline{\cb}^e(\ub)$.

\section{Model validation and estimation}
\label{sect:simest}
\subsection{Comparison with ergometer datasets}

In this section we calibrate the proposed model based on the physiological data presented in Fig.~\ref{fig:compfig}, corresponding to one \emph{single} representative volunteer breathing an atmosphere free of isoprene (i.e., we set $\cinh \equiv 0$ in the sequel). It will turn out that the model appears to be flexible enough to capture the isoprene profiles in exhaled breath generally observed during moderate workload ergometer challenges as conducted in~\citep{King2009}. Moreover, our formulation provides a preliminary basis for estimating some of the unspecified parameters $p_j \in \{\prl,\prm,\ml,\mm,\tau,\vper \}$ from the knowledge of measured breath concentrations $y$.
More specifically, our aim is to (at least partially) determine the \emph{subject-dependent} parameter vector 
\[\tb=(\prl,\prm,\ml,\mm,\tau,\vper)\]
as well as the nominal endogenous steady state levels $\cb_0=\cb(t_0)$ by solving the ordinary least squares problem
\begin{equation}\label{eq:ls}\underset{\tb,\cb_0}{\mathrm{argmin}} \sum_{i=0}^n \big(y_i-\calv(t_i)\big)^2,\end{equation}
subject to the constraints
\begin{equation}\label{eq:const}\left\{\begin{array}{ll}
\gb(\ub_0,\tb,\cb_0)=\nulb & \textrm{(steady state)}\\
\tb,\cb_0 \geq \nulb & \textrm{(positivity)}\\
\underline{\cb}^e_4(\ub_0,\tb)=25\;\textrm{nmol/l} & \textrm{(exposure steady state).} \end{array} \right.
\end{equation}
Here, $\gb$ is the right-hand side of the ODE system~\eqref{eq:alv}--\eqref{eq:per} (see also~\eqref{eq:sysc}) and $y_i=C_{\mathrm{meas},i}$ is the measured end-tidal isoprene concentration at time instant $t_i$ ($t_0=0$). The solution point will be denoted by $(\tb^*,\cb_0^*)$. For perspective, the last constraint has been introduced in order to account for additional information regarding the biotransformation of isoprene available on the basis of toxicological inhalation studies~\citep{f1996}. As has been demonstrated there for an ensemble of four normal healthy test subjects, isoprene concentrations in a closed rebreathing chamber of fixed volume will plateau at a level of approximately 600~ppb after about 2~hours of quiet tidal breathing at rest, irrespective of the initial amount of isoprene present in the system. The extracted parameters will be adjusted to automatically meet this boundary condition, thereby maintaining consistency with the aforementioned experimental findings.

For simulation purposes the measured physiological functions $\qalv$ and $\qc$ were converted to input function handles $\ub$ by applying a local smoothing procedure to the associated data and interpolating the resulting profiles with splines. 
Tissue volumes and partition coefficients are as in Table~\ref{table:param}. In particular, while the peripheral compartment so far has been treated as an abstract control volume without particular reference to any specific tissue group, for identifiability reasons we now set $\lper=0.5$, which corresponds to the in vitro blood:tissue partition coefficient for muscle~\citep{f1996}. Note, however, that this choice is rather arbitrary, cf. Remark~\ref{rem:unid}.\\

\begin{figure}[H]
\centering
\begin{tabular}{c}
\includegraphics[width=9cm]{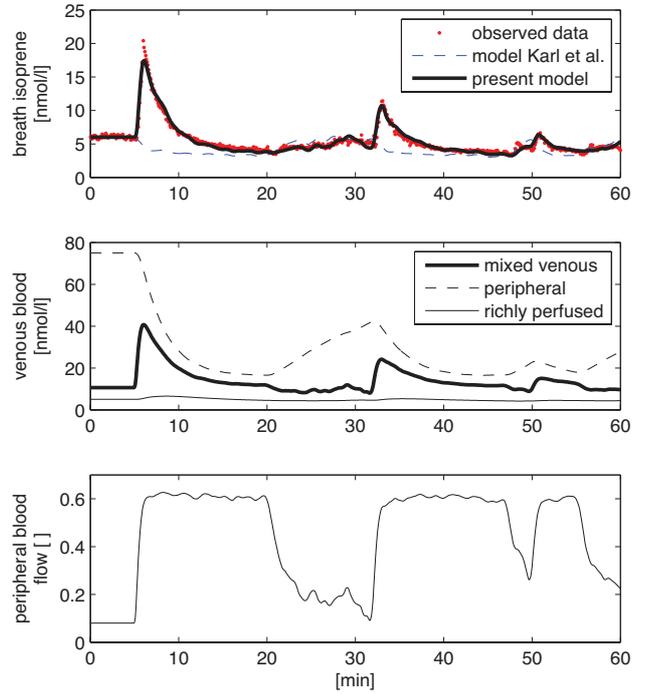}
\end{tabular}
\vspace{-0.5cm}
\caption{First panel: simulation of end-tidal isoprene behavior during exercise conditions, cf. Fig.~\ref{fig:compfig}. Second panel: predicted concentrations in mixed venous blood ($\cven$) and venous blood returning from the peripheral ($\lper\cper$) and richly perfused tissue groups ($\lrpt\crpt$). Third panel: predicted profile of fractional peripheral blood flow $\qper$ according to Equation~\eqref{eq:qper}.}\label{fig:fit}
\end{figure}

The above minimization problem~\eqref{eq:ls} was solved by implementing a multiple shooting routine~\citep{bock1987} in \textit{Matlab}. This iterative method can be seen as a generalization of the standard Gauss--Newton algorithm, designed to avoid divergence issues of the latter due to large residuals. For further details as well as convergence and stability properties we refer to~\citep{bock1981,pfeifer2007}. The necessary derivatives of the trajectories with respect to $\tb$ and $\cb_0$ were computed by simultaneously solving the associated variational equations~\citep{wannerbook}. Convergence was assumed to be achieved when the maximum componentwise relative change between two successive iterations was less than 0.1\%. Fig.~\ref{fig:fit} summarizes the results of these calculations. Fitted parameter values and initial conditions are given in Table~\ref{table:fit}. \\

All estimated quantities for the test subject under scrutiny take values in a physiologically plausible range. According to Equations~\eqref{eq:cart} and~\eqref{eq:cven}, arterial and mixed venous blood concentrations at the start of the experiment are estimated as $\cart(0)=4.5$~nmol/l and $\cven(0)=10.6$~nmol/l, respectively, which is in direct agreement with available data from the literature (cf.~Table~\ref{table:param}). Total endogenous production equals approximately 125~nmol/min, which is comparable to previous predictions ranging from 2.5~to~5.7~nmol/min/kg bodyweight~\citep{hartmann1990,f1996}. Moreover, the estimated value for $\vper$ is close to experimentally measured thigh muscle volumes (see~\citep{tothill2002} for instance). 

For the sake of comparison, in Fig.~\ref{fig:fit} we also show the outcome of the model by Karl~et~al. subjected to the time courses of $\qalv$ and $\qc$ as above (assuming the same end-tidal steady state value of 6~nmol/l at rest). As has been indicated in Section~\ref{sect:res}, the associated predictions result in a poor representation of the observed data. 

The local identifiability of the extracted estimates in Table~\ref{table:fit} was investigated by checking the non-singularity of the information matrix $Q:=S^T S$, where $S$ is the sensitivity function matrix having rows
\begin{equation}
S_{i,-}:=\begin{pmatrix}\frac{\partial y(t_{i-1},\tb^*,\cb_0^*)}{\partial \tb} & \frac{\partial y(t_{i-1},\tb^*,\cb_0^*)}{\partial \cb_0}\end{pmatrix}.
\end{equation}
More specifically, we adopted the standard \emph{numerical} rank criterion
\begin{equation}
\mathrm{rank}\,Q=\max\{k;\;\sigma_k >\varepsilon \norm{Q}_{\infty}\},
\end{equation}
where $\sigma_1\geq \sigma_2 \geq \ldots \geq 0$ are the singular values of $Q$ and $\varepsilon = 10^{-8}$ denotes the maximum relative error of the calculated sensitivities~\citep{golubbook}. Accordingly, we find that $Q$ has full rank, suggesting that all estimated quantities are practically identifiable~\citep{cobelli1980}. However, some degree of ill-conditioning is present as can be concluded from calculating the approximate posterior correlation matrix $R$ defined by
\begin{equation}
R_{i,j}:=Q^{-1}_{i,j}\big(Q^{-1}_{i,i}Q^{-1}_{j,j}\big)^{-\frac{1}{2}} \in [-1,1].
\end{equation}
The entry $R_{i,j}$ quantifies the degree of interplay between the $i$th and $j$th parameter (initial condition) under scrutiny. 

A value of $R_{i,j}$ near $+1$ or $-1$ indicates that it may be difficult to estimate both parameters separately, as changes in the model output caused by perturbing one of these parameters can nearly be compensated by an appropriate perturbation of the other~\citep{jac1990,seberbook,rodriguez2006}. The highest correlation is achieved for the pair $(\prl,\ml)$, with an associated value of~0.995. This indicates a poor estimability of the above-mentioned two parameters if only the breath isoprene dynamics in Fig.~\ref{fig:fit} are taken into account. However, the constraints in~\eqref{eq:const} provide additional information on $\prl$ and $\ml$ that will prove sufficient for guaranteeing the extraction of reliable estimates.
Alternatively, such identifiability issues might also be circumvented by designing multi-experimental regimes guaranteeing a sufficiently large and independent influence of all parameters under scrutiny (for instance, by complementing ergometer challenges with closed chamber rebreathing protocols as indicated above). The absolute value of all other pairwise correlations is below~0.9.\\

\begin{figure}[H]
\centering
\begin{tabular}{c}
\includegraphics[width=8cm]{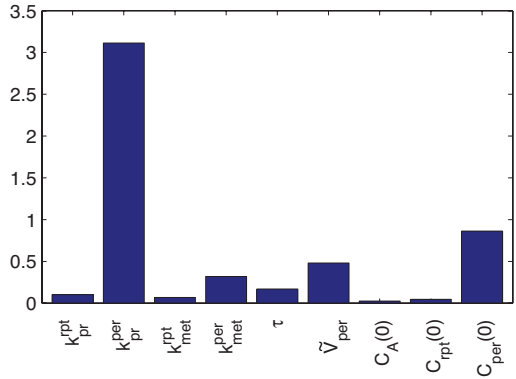}
\end{tabular}
\vspace{-0.5cm}
\caption{Squared $L_2$-norm of the \emph{normalized} model sensitivities (cf. Equation~\eqref{eq:sens}) with respect to the fitted parameters in Table~\ref{table:fit}.}\label{fig:sens}
\end{figure}

A ranking of the fitted parameters and initial conditions with respect to their impact on the model output can be obtained by numerically approximating the squared $L_2$-norm of the \emph{normalized} sensitivities, viz.,
\begin{equation}\label{eq:sens}
\varsigma(p_j):=\int\limits_{t_0}^{t_n}\lk\frac{\partial y(t,\tb^*,\cb_0^*)}{\partial p_j}\frac{p_j^*}{\max_s \abs{y(s)}}\rk^2 \mathrm{d} t,
\end{equation}
and similarly for the components of $\cb_0$. A graphical comparison of these sensitivity indices is given in Fig.~\ref{fig:sens}, revealing a strong influence of $\prm$, $\cper(0)$ and $\vper$ on the predicted breath isoprene profile. This is intuitively reasonable as these quantities govern the shape of the observed isoprene peak during exercise. 

Contrarily, only minor effects are seen when varying the (poorly determined) parameters $\prl$ and $\ml$. In fact, it should be pointed out that production and metabolization in the richly perfused tissue group are not needed for producing a satisfactory fit of the data given in Fig.~\ref{fig:fit}. However, we refrained from generally eliminating these variables as they play a major role in isoprene distribution during resting conditions (when blood flow is directed mainly to the richly perfused tissue compartment). In particular, they ensure the consistency of the model with closed chamber rebreathing scenarios as discussed before.
From the ensemble of fixed model parameters, the most influential quantities (having a sensitivity index value greater than~0.25 according to Eq.~\eqref{eq:sens}) are the maximum fractional perfusion to peripheral tissue ($\varsigma(\qper^{\mathrm{max}})=0.88$), the partition coefficient between blood and peripheral tissue ($\varsigma(\lper)=0.6$) and the blood:gas partition coefficient ($\varsigma(\hen)=0.26$). These variables should be given special attention when applying the proposed model to a larger study population as they require a careful assessment with respect to inter-individual variations. \\

In order to give some insight into the information content of the extracted parameter values, approximate standard errors were constructed by employing a variant of \emph{residual bootstrapping}~\citep[Sect. 2.3.5]{dogan2007,huetbook}. For an excellent overview of resampling techniques in general the interested reader is referred to~\citep{shaobook}, while a recent comparison between standard asymptotic theory and bootstrapping for uncertainty quantification in inverse problems can be found in~\citep{banks2010}. 

In particular, this method allows for taking into account autocorrelations detected among the model residuals 
\begin{equation}\label{eq:res}
r_i:=y_i-y(t_i,\tb^*,\cb_0^*),\quad i=0,\ldots,n.
\end{equation}
Such autocorrelation patterns can be seen as a general feature of dense time course, ventilation-related data streams~\citep{liang1996} and neglecting their presence typically tends to distort variance assessments of least squares estimates derived from conventional covariance matrix approximations~\citep{seberbook,davidianbook}. Adopting the general procedure suggested by~\citet{dogan2007}, we first use standard techniques from time series analysis (see, e.g.,~\citep{boxbook}) to model the interdependence between the $r_i$ via an autoregressive process of order two, viz.,
\begin{equation}\label{eq:ar}
r_i=\alpha r_{i-1}+ \beta r_{i-2} + \tilde{r}_i.
\end{equation}
Plots of the resulting $\tilde{r}_i$ versus time clearly exhibit random patterns, thereby suggesting that the former can be treated as independent and homoscedastic realizations of the underlying error process. Furthermore, a Ljung--Box portmanteau test~\citep{ljung1978} confirmed the lack of statistically significant autocorrelations. We can hence conclude that the error terms $\tilde{r}_i$ are interchangeable.

Consequently, a single bootstrap dataset $\yb^b:=\big(y_0^b,\ldots,y_n^b\big)$ may be generated by the following procedure: we draw $n-1$ samples from a uniform discrete distribution over the set $\{\tilde{r}_i;\,i=0,\ldots,n\}$. The results are combined to yield a vector $\big(\tilde{r}_2^b,\ldots,\tilde{r}_n^b\big)$, from which $\yb^b$ is obtained via Equations~\eqref{eq:ar} and~\eqref{eq:res} (we set $r_i^b:=r_i$ for $i=0,1$). This resampled dataset is then plugged into the minimization procedure~\eqref{eq:ls} to arrive at new estimates $\big(\tb_{\phantom{0}}^{*,b,}\cb_0^{*,b}\big)$. Repeating the above step $B$ times generates a population of $B$ fits for each component of $\tb$ and $\cb_0$, reflecting the sensitivity of these estimates with respect to the given data. Approximate standard errors might then be computed from the empirical variances associated with these populations. Here, we use $B=100$.\\

The variation coefficients in Table~\ref{table:fit} suggest that under the constraints imposed in~\eqref{eq:const} all unknown parameters and initial conditions might be determined from the individual breath concentration data in Fig.~\ref{fig:compfig} with reasonable accuracy. While this confirms that inference on endogenous isoprene kinetics by virtue of exhaled breath measurements is potentially feasible, it must be emphasized that the extracted values are clearly model-dependent. In particular, additional modeling efforts investigating a more refined compartmentalization and description of perfusion patterns as in Equation~\eqref{eq:qper} will be imperative before such estimates can become practically relevant. Moreover, further experimental evidence needs to be gathered with respect to (fixed) physiological parameters that are known to drastically affect the model output. Sensitivity and identifiability methodologies as indicated above can guide these tasks (see also~\citep{brun2002,arias2009,hengl2007}). In this sense, the preceding analysis should merely be seen as a preliminary proof of concept, that primarily aims at proposing a novel qualitative description of the normal physiological flow of isoprene rather than at drawing further quantitative conclusions with respect to the indicated estimates. 

\begin{table}[H]
\centering \footnotesize
\begin{tabular}{|lccc|}\hline
{\large\strut}Variable&Symbol&Fitted value (units) & CV\\ \hline \hline 
{\large\strut}Production rpt &$\prl$ & 20.8 (nmol/min) & 16\\
{\large\strut}Production periphery &$\prm$ & 104.5 (nmol/min) & 3\\
{\large\strut}Metabolism rate rpt &$\ml$ & 3.6 (l/min) & 8\\
{\large\strut}Metabolism rate periphery &$\mm$ & 0.96 (l/min) & 7\\
{\large\strut}Constant Eq.~\eqref{eq:qper} &$\tau$ & 2.1 & 12\\
{\large\strut}Tissue volume periphery &$\vper$ & 9.2 (l) & 8\\
{\large\strut}Initial concentration alveoli &$\calv(0)$ & 6 (nmol/l) & 3\\
{\large\strut}Initial concentration rpt &$\crpt(0)$ & 12.5 (nmol/l) & 6\\
{\large\strut}Initial concentration periphery &$\cper(0)$ & 150 (nmol/l) & 5\\
\hline
\end{tabular}
\caption[small]{Decisive model parameters resulting from the fit in Fig.~\ref{fig:fit}. The corresponding variation coefficients (CV, in~\%) were obtained by calculating bootstrap standard errors from the repeated fits of $B=100$ resampled datasets.}\label{table:fit}
\end{table}

Moreover, we again stress the fact that the fitting procedure above has been carried out for one single representative volunteer only, inasmuch as our major goal was to demonstrate the principal explanatory power of the proposed model for capturing the presented breath isoprene behavior. 
The population spread of the fitted parameters within the larger study cohort investigated by~\citet{King2009} might be assessed by a Bayesian (see~\citep{moerk2009} for instance) or mixed effects approach~\citep{kuhn2005}, which, however would be beyond the scope of this paper.\\

\begin{rmk}
For the sake of completeness, we briefly note that a formal description of the experimental situation during the one-legged ergometer trials as in Section~\ref{sect:res} can be obtained by simply augmenting the model with a copy of Equation~\eqref{eq:per}. For symmetry reasons, each of these two peripheral compartments (interpreted as left and right leg) might then be assigned 50\% of the volume $\vper$, nominal fractional blood flow $\qper$, production $\prm$ and metabolization rate $\mm$ as given in Table~\ref{table:fit} and Table~\ref{table:param} (note that the initial steady state concentrations remain unchanged). Consequently, by alternately distributing increased fractional perfusion during the individual exercise segments to either one of these compartments, a good qualitative agreement with the data shown in Fig.~\ref{fig:compfig2} can be achieved.   
\end{rmk}

\subsection{Physiological interpretation}

The second panel in Fig.~\ref{fig:fit} clearly reveals the physiological mechanism underlying the peak shaped dynamics of breath isoprene concentrations in response to constant load exercise. During rest, the peripheral compartment is characterized by high isoprene concentrations resulting from extrahepatic production according to $\prm$. However, due to the minute fractional blood flow $\qper^{\mathrm{rest}}$ to these tissues, \emph{mixed} venous concentrations are mainly governed by the lower values in venous blood from the rpt group. As soon as fractional perfusion in the periphery increases as a result of exercise hyperemia, mixed venous concentrations become dominated by peripheral venous return. The isoprene peak visible in mixed venous blood and breath is an immediate consequence of this transition. Subsequently, a depletion of the peripheral tissue compartment and hence a decline in mixed venous blood concentration can be observed. As a matter of fact, if $\qalv^{\mathrm{work}}$ and $\qc^{\mathrm{work}}$ are maintained at a roughly fixed level reflecting some constant workload, the compartmental concentrations will approach a new steady state $\cb^e(\ub^{\mathrm{work}})$, which is attained after about 15~minutes of pedaling, cf. Section~\ref{sect:res}. When the workload is stopped, perfusion will be redistributed according to the compartmental shares at rest and the peripheral isoprene buffer will be replenished. If exercise is continued before this process is completed, the corresponding isoprene peak will be lower than at the start of the first exercise segment, despite a similar response of ventilation and perfusion. This clarifies the wash-out behavior discernible in repeated workload segments. In the special situation of Fig.~\ref{fig:fit}, starting from the final state at $t=20$~min (using the fitted parameter values in Table~\ref{table:fit} and applying the physiological inputs $\ub_0$ corresponding to resting conditions) the time required until all compartmental concentrations are within 1\% of their initial level $c_{0,i}$ can be simulated as approximately 58~min. This is consistent with experimental observations~\citep{King2009}. \\

In other words, according to the preceding rationale the major part of breath isoprene variability during ergometer challenges can be attributed to varying fractional contributions of distinct compartmental levels to the mixed venous blood concentration $\cven$. The aforementioned reasoning compares favorably with the fact, that peripheral venous blood concentrations (median 30~nmol/l; range 15-70~nmol/l~\citep{cailleux1992}) appear to be significantly higher than mixed venous ones (median 9~nmol/l; range 0.5-24~nmol/l~\citep{miekischblood}). In particular, note that with the present model the observed isoprene dynamics can be explained assuming \emph{constant} endogenous production rates, which agrees with the intuitive perception of isoprene synthesis as a slowly varying process. In this sense, the aforementioned putative mechanism optimally respects a wide spectrum of fundamental phenomenological as well as physiological boundary conditions. 
From a practical point of view, the intimate ties between compartmental hemodynamics and endogenous isoprene flow put forward by the previous analysis might render breath isoprene as a promising new parameter for studying vascular control and the redistribution of blood flow during exercise.\\

\begin{rmk}
A word is in order regarding the necessity of introducing a hypothetical production rate $\prm$ for ensuring the formation of a systemic isoprene pool.
To this end, consider an arbitrary non-producing and non-metabolizing body compartment which may essentially be characterized by a mass balance equation of the form \eqref{eq:per}, with $\prm$ and $\mm$ set to zero (the index ``$\mathrm{per}$'' is kept merely for notational convenience). As steady state conditions can be assumed to hold during rest (see Section~\ref{sect:res}), the initial venous concentration $\cper(0)\lper$ of isoprene associated with this compartment will be equal to the incoming arterial concentration $\cart(0)$ of the compound. Adopting the above notation we thus find that
\begin{equation}
\cper(0)\lper=\cart(0) \leq \cven(0).
\end{equation}
The last inequality is a consequence of the algebraic steady state relation associated with the alveolar compartment (cf. Equation~\eqref{eq:farhi}, which is a standard mass balance equation for gas exchange in the lung).
Hence, when switching to an increased fractional perfusion of such body compartments as a result of exercise, the \emph{mixed} venous return will become enriched with blood having isoprene concentrations close to the previous arterial level during rest. In other words, $\cven$ will fall rather than rise. Using this simple but general rationale it is clear why previous models of isoprene pharmacokinetics such as in~\citep{f1996} fail to reproduce the peak-shaped behavior of breath isoprene during exercise, even if differential blood flow is taken into account.
\end{rmk}

Regarding further model validation, the experimental outcome associated with the one-legged ergometer regimes presented in Section~\ref{sect:res} appears to furnish the fact that any quantitative formulation neglecting a peripheral release mechanism of isoprene will be an inappropriate physiological description of the prevailing isoprene dynamics during exercise. However, further biochemical and physiological studies will have to be conducted in order to pinpoint the exact origin of this effect. Apart from the line of argumentation presented above, alternative isoprene sources might comprise
\begin{itemize}
\item[(a)] an exercise-induced, time-varying production in contracting muscle (possibly due to rapid switches in cellular metabolism) 
\item[(b)] a change of diffusion capacities in peripheral tissue (reflected, for instance, by an abrupt increase of $\lper$, cf. Equation~\eqref{eq:cven}).
\end{itemize}
However, note that while (a)~is not consistent with our current understanding of the isoprene synthetic pathway and does not provide a natural explanation for the distinct peak heights observed in repeated workload regimes, (b)~appears questionable due to the fact that such a transition is likely to influence both isoprene and butane kinetics in a similar way. This contradicts experimental evidence (see the discussion in Section~\ref{sect:res} and~\citep{King2010GC}).  

\begin{rmk}\label{rem:unid}
It should be mentioned that a more precise specification of the peripheral tissue compartment on the basis of estimated volumes and partition coefficients could not be achieved. For instance, choosing $\lper=1/82$ (which is the proposed blood:tissue partition coefficient for fat~\citep{f1996}) and setting $\vper = 0.23$~l as well as $\cper(0) = 6147$~nmol/l in Table~\ref{table:fit} yields a fit of similar quality as in Fig.~\ref{fig:fit}. With these modifications in mind, contrary to the previous interpretation as muscle tissue, the peripheral compartment might hence also be viewed as a small isoprene buffer volume characterized by a high lipid content (such as for instance the endothelial layer lining the vascular walls). This lack of joint estimability of $\lper$ and $\vper$ within the present experimental setting is also reflected by a high degree of collinearity between the associated sensitivities $\partial y/\partial \lper$ and $\partial y/\partial \vper$, respectively.
\end{rmk}

\section{Conclusion}

This paper is devoted to the development of a first mechanistic description of isoprene evolution in different tissue compartments of the human body by simulating the behavior of breath isoprene output during several short-term exercise protocols. In Section~\ref{sect:res} various lines of supportive experimental evidence for an extrahepatic tissue source of isoprene have been presented. These findings have led us to a simple kinetic model that is expected to aid further investigations regarding the exhalation, storage, transport and biotransformation processes associated with this important compound. \\

The emphasis of this work has been laid on deriving a sound mathematical formulation flexible enough to cover a wide spectrum of possible isoprene behavior in end-tidal breath, while simultaneously maintaining consistency with earlier experimental findings as well as physiological plausibility of the involved parameters. Depending on the specific field of application, necessary model refinements might include the incorporation of a multi-compartment lung for mapping ventilation--perfusion mismatch or changes in diffusion capacity, as well as a less coarse partition of the systemic tissue groups, similar as in~\citep{f1996,melnick2000}. The statistical significance of these generalizations might then be assessed, e.g., by employing residual-based comparison techniques for nested models as described in~\citep{banksbook,banks1990}. However, at the current stage of research and given the limited data on the dynamic behavior of breath isoprene throughout a broader spectrum of experimental scenarios, it is preferable to maintain a compartmentalization and parameterization as parsimonious as possible.\\

On-line determinations of dynamic VOC concentration profiles in exhaled breath combined with adequate kinetic modeling is a promising field of research, still in its
infancy. From a methodological point of view, this work demonstrates that such dynamic patterns reflect fundamental physiological changes and can potentially be used for exploring the fate of volatile species in the human body. Generally, it should also be emphasized that a reliable quantification of relevant substance-specific characteristics of endogenous trace gases (such as production and metabolism) from breath data might yield novel diagnostic or therapeutic indicators that are complementary to those gained by employing more invasive methods. In this sense, we hope that the present contribution will help to consolidate the potential role of breath gas analysis in biomonitoring and will also stimulate future efforts to establish mathematical modeling as a core technique in VOC research.

\section*{Acknowledgements}
We are indebted to the reviewers for several helpful suggestions.
Julian King is a recipient of a DOC fellowship at the Breath Research Institute of the Austrian Academy of Sciences.
The research leading to these results has received funding from the
European Community’s Seventh Framework Programme~(FP7/2007-13) under
grant agreement No.~217967. We appreciate funding from the Austrian
Federal Ministry for Transport, Innovation and Technology (BMVIT/BMWA,
Project~818803, KIRAS). Gerald Teschl and Julian King acknowledge support from the Austrian Science Fund (FWF) under Grant No.~Y330. 
We greatly appreciate the generous support of the government of Vorarlberg
and its governor Landeshauptmann Dr. Herbert Sausgruber. 


\appendix
\section{Classical inert gas elimination theory}\label{app:farhi}

Adopting the nomenclature in Table~\ref{table:param}, the basic equation for modeling pulmonary exchange of blood-borne inert gases using one single lung compartment
is a mass balance equation of the form (see, e.g., 
\citep{bkst2007})
\begin{align}
V_A \frac{ \di C_{A} }{\di t} =\dot{V}_{A} ( C_{I} -C_{A}) +  \dot{Q}_{c} (C_{\bar v} -C_{a}), \label{eq:ap1} 
\end{align}   
where $C_{X}$ denotes the trace gas concentration in a region $X$ averaged over a period $\Delta t$,   
i.e., 
\begin{equation}C_{X}(t) = 1/ \Delta t \!\! \int \limits_{t-\Delta t/2}^{t+\Delta t/2} \!\!\hat C_{X}(s) \di s.
\end{equation}

From Equation~\eqref{eq:ap1}, by assuming steady state conditions $\di C_{A}/\di t  =0$ as well as $C_{I}=0$ (i.e., no trace gas is inspired) and by substituting Henry's law $C_{a} = \hen \, C_{A}$ we derive the familiar equation due to~\citet{farhi1967},
\begin{align}\label{eq:farhi}
\cmeas = C_{A} =  \frac{C_{\bar v}}{\hen + \frac{\dot{V}_{A}}{\dot{Q}_{c}}}. 
\end{align} 
Here, the quotient $\dot{V}_{A}/\dot{Q}_{c}$ is called ventilation--perfusion ratio, whereas $\hen$ denotes the substance-specific and temperature-dependent blood:gas partition coefficient.

\section{Some fundamental model properties}\label{sect:apriori}
Here we shall briefly recall some general properties of the proposed model that necessarily must be satisfied in any valid description of concentration dynamics. Firstly, note that Equations~\eqref{eq:alv}--\eqref{eq:per} can be written as a time-varying, linear inhomogeneous system
\begin{equation}\label{eq:sysc}
\dot{\cb}=A(\ub,\tb)\cb+\bb(\ub,\tb)=:\gb(\ub,\tb,\cb)
\end{equation}
in the state variable $\cb:=(\calv,\crpt,\cper)^T$, which is dependent on a constant parameter vector $\tb$ as well as on a vector
$\ub:=(\qalv,\qc,\cinh)$ lumping together all measurable external inputs.

Non-negativity of the trajectories associated with~\eqref{eq:sysc} for non-negative initial conditions easily follows from the fact that the system is cooperative. Moreover, by considering the dynamics of the total amount of isoprene $m:=\sum_i \tilde{V}_i c_i \geq 0$, viz.,
\begin{multline}\label{eq:massderiv}
\dot{m}=\prm+\prl-\mm\lper\cper-\ml\lrpt\crpt+\\\qalv(\cinh-\calv),
\end{multline}
it can readily be verified that the trajectories are bounded from above if either $\qalv>0$ or if at least one of the two metabolic rates $\ml$ or $\mm$ is strictly positive. Furthermore, it can be proven that under physiological steady state conditions, i.e., for constant $\ub$, the time-invariant matrix $A$ will be Hurwitz if
\[\det(A) = \qalv \vartheta_1 + \ml\vartheta_2 + \mm\vartheta_3 + \ml\mm\vartheta_4  \ne 0,\quad \vartheta_i < 0,\]
cf.~\citep[Prop.~2]{King2010a}. Hence, except for the degenerate case $\qalv=\ml=\mm=0$ (which, as can be seen from~\eqref{eq:massderiv}, necessarily results in divergent trajectories if one of the two production rates is strictly positive) the compartmental concentrations can be guaranteed to approach a globally asymptotically stable equilibrium $\cb^e(\ub):=-A^{-1} \bb$ once the inputs $\ub$ affecting the system are fixed.

\newpage

\section{Nomenclature}
\setcounter{table}{0}
\begin{table}[H]
\centering 
\footnotesize
\begin{tabular}{|lcr|}\hline
 {\large\strut}Parameter&Symbol&Nominal value (units)\\ \hline \hline
{\large\strut}\textit{Concentrations} & & \\ 
{\large\strut}  alveoli &$\calv$ & 4 (nmol/l)$^a$\\
{\large\strut}  end-capillary &$C_{\mathrm{c'}}$ & \\
{\large\strut}  arterial &$\cart$ & 5.7 (nmol/l)$^b$\\
{\large\strut}  mixed-venous &$\cven$ & 9 (nmol/l)$^b$\\
{\large\strut}  richly perfused tissue (rpt) &$\crpt$ & \\
{\large\strut}  peripheral tissue &$\cper$ & \\
{\large\strut}  inhaled (ambient) &$\cinh$ & 0 (nmol/l) \\
{\large\strut}\textit{Compartment volumes} & & \\
{\large\strut}  alveoli &$V_{\mathrm{A}}$ & 4.1 (l)$^c$\\
{\large\strut}  end-capillary &$V_{\mathrm{c'}}$ & 0.15 (l)$^d$\\
{\large\strut}  richly perfused (rpt) &$V_{\mathrm{rpt}}$ & 13.25 (l)$^e$\\
{\large\strut}  blood rpt &$V_{\mathrm{rpt,b}}$ & 1.97 (l)$^e$\\
{\large\strut}  peripheral tissue &$V_{\mathrm{per}}$ & \\
{\large\strut}  blood peripheral tissue &$V_{\mathrm{per,b}}$ & \\
{\large\strut}  ambient &$\vinh$ & \\
{\large\strut}\textit{Fractional blood flows} & & \\
{\large\strut}  periphery (both legs)&$\qper$ & \\
{\large\strut}  maximal &$\qper^\mathrm{max}$ & 0.7$^{f}$\\
{\large\strut}  nominal (rest) &$\qper^\mathrm{rest}$ & 0.08$^g$,\;0.14$^h$\\
{\large\strut}  constant Eq.~\eqref{eq:qper} &$\tau$ & \\
{\large\strut}\textit{Partition coefficients} & & \\
{\large\strut}  blood:air &$\hen$ & 0.75$^{i,j}$\\
{\large\strut}  blood:rpt &$\lrpt$ & 0.4$^j$ \\
{\large\strut}  blood:peripheral tissue &$\lper$ & 0.5 (muscle)$^j$; 0.012 (fat)$^j$\\
{\large\strut}\textit{Rate constants} & & \\
{\large\strut}  hepatic metabolic rate &$\ml$ & \\
{\large\strut}  extrahepatic metabolic rate &$\mm$ & \\
{\large\strut}  production rpt &$\prl$ & \\
{\large\strut}  production peripheral tissue &$\prm$ & \\
\hline
\end{tabular}
\caption{Basic model parameters and reference values for normal subjects during rest; $^a$\citep{kushch2008}; $^b$mechanically ventilated patients in~\citep{miekischblood}; $^c$\citep{moerk2006}; $^d$\citep{pulmcirc}; $^e$comprising viscera, brain and connective muscles according to Table~8.2 in ~\citep{ottesen2004}; $^f$corresponding to 450~kpm/min or approx. 75~W according to Fig.~6 in~\citep{sullivan1989}; $^g$\citep{johnson2007}; $^h$obtained by $\qper^\mathrm{rest}=2~(\textrm{single leg blood flow}/\textrm{cardiac output})$ according to Table~1 in~\citep{sullivan1989}; $^i$\citep{kpm01}; $^j$\citep{f1996}.}\label{table:param}
\end{table}

\bibliographystyle{elsart-harv}
\bibliography{IsopreneBib}

\end{document}